\documentclass[aps,prl,superscriptaddress,twocolumn,preprintnumbers,floatfix]{revtex4-2}
\pdfoutput=1
\usepackage[utf8]{inputenc}
\usepackage{tikz-cd}
\usepackage{color}
\usepackage{dsfont}
\usepackage{bm}
\usepackage{mathtools}
\usepackage{amsthm}
\usepackage{amsfonts,amssymb,amsmath}
\usepackage[hidelinks]{hyperref}
\usepackage{physics}
\usepackage{multirow}
\usepackage{graphicx}
\usepackage{ulem}
 \newtheorem{theorem}{Theorem}
  \newtheorem{lemma}{Lemma}

\usepackage{verbatim}

\begin{document}

\title{
Critically Slow Hilbert-Space Ergodicity in Quantum Morphic Drives
}

\author{Sa\'ul Pilatowsky-Cameo}
\affiliation{Center for Theoretical Physics, Massachusetts Institute of Technology, Cambridge, Massachusetts 02139, USA}
\author{Soonwon Choi}
\affiliation{Center for Theoretical Physics, Massachusetts Institute of Technology, Cambridge, Massachusetts 02139, USA}
\author{Wen Wei Ho}
\affiliation{Department of Physics, National University of Singapore, Singapore 117551}
\affiliation{Centre for Quantum Technologies, National University of Singapore, 3 Science Drive 2, Singapore 117543}
\preprint{MIT-CTP/5841}
\begin{abstract}
The maximum entropy principle is foundational for statistical analyses of complex dynamics. This principle has been  challenged by the findings of a previous work [\href{https://doi.org/10.1103/PhysRevX.7.031034}{Phys.~Rev.~X {\bf 7}, 031034 (2017)}], where it was argued that a quantum system driven in time by a certain aperiodic  sequence without any explicit symmetries, 
dubbed the Thue-Morse drive, 
gives rise to emergent nonergodic steady states which are underpinned by effective conserved quantities.
Here, we resolve this apparent tension. 
We rigorously prove that the Thue-Morse drive achieves a very strong notion of quantum ergodicity in the long-time limit: The time evolution of any initial state  uniformly visits every corner of its Hilbert space. On the other hand, we find the dynamics also approximates a Floquet drive for arbitrarily long albeit finite periods of time 
with no characteristic timescale, resulting in a scale-free ergodic dynamics we call critically slow complete Hilbert-space ergodicity. 
Furthermore,  numerical studies reveal that critically slow complete Hilbert-space ergodicity  is not specific to the Thue-Morse drive and is, in fact, exhibited by many other aperiodic  drives derived from morphic sequences, i.e., words derived from repeatedly applying substitution rules on basic characters. 
Our work presents a new class of  dynamics in time-dependent quantum systems where full ergodicity is eventually attained, but only after astronomically long times.
\end{abstract}
\maketitle

The maximum entropy principle (MEP) is a powerful tenet in physics, guiding predictions about the limiting behavior of a complex system without 
the need for a full, microscopic solution~\cite{Jaynes1957}. It posits that at equilibrium, a system is well described by a state which maximizes a suitable notion of   entropy, constrained only by   globally conserved quantities. For example, the MEP predicts that the steady state achieved in thermalization is a Gibbs state, characterized solely by a few macroscopic properties like temperature and chemical potential.

 The MEP has also guided recent studies of quantum ergodicity  concerning the temporal ensemble of wave functions of a closed quantum system \cite{Roberts2017,Pilatowsky2023,Pilatowsky2024,Shaw2024,Mark2024,Ghosh2024a,Logaric2024,Liu2025}.
These works inquire about the limiting  {\it distribution} 
formed by the collection of pure states arising in evolution and go beyond traditional notions of quantum ergodicity involving statistics of energy eigenvalues or eigenstates 
\cite{DAlessio2016}. For example, when dynamics possesses a time-translation symmetry (e.g., generated by a time-independent or periodic Hamiltonian),  the MEP predicts that the {\it random phase ensemble} emerges --- a collection of states wherein energy eigenstates are superposed with random phases \cite{Shaw2024,Mark2024}.
In contrast, in the absence of time-translation and  other internal symmetries,   the MEP suggests 
the emergence of the
maximally ergodic, uniform (Haar) distribution, a phenomenon dubbed {\it complete Hilbert-space ergodicity} (CHSE)~\cite{Pilatowsky2023, Pilatowsky2024}. Indeed, Ref.~\cite{Pilatowsky2023} rigorously showed that CHSE  arises in the discrete time-quasiperiodic  Fibonacci drive, which is derived from the Fibonacci word~
\cite{Nandy2018,Sayak2019,Dumitrescu2018,Dumitrescu2022,Das2023,Ghosh2024,Tiwari2024}. 
More generally, the same universal maximally ergodic behavior is expected to hold for generic quantum systems driven  {\it aperiodically} in time, which do not feature any explicit conserved quantities. 

\begin{figure}[t]
    \centering
    \includegraphics[width=0.95\columnwidth]{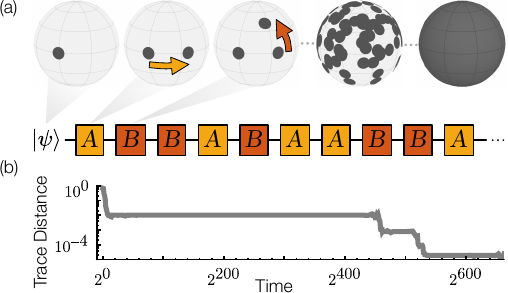}
    \caption{(a) The Thue-Morse drive involves applying a pair of unitaries $A$ and $B$ according to the Thue-Morse word. We ask if, under such dynamics, a quantum state uniformly covers the Hilbert space  over time. (b) Trace distance between the temporal distribution [dotted spheres in (a)] and the  uniform distribution (dark sphere)   for a representative initial state. One sees a nonsmooth decay  involving many intervening long-lived plateaus, 
    each lasting for astronomically long times. 
     It is not clear whether the trace distance keeps decreasing (CHSE) or eventually saturates at a nonzero value. 
    }
    \label{fig:01}
\end{figure}

However, a previous study \cite{Nandy2017} of the equilibration dynamics in the quantum Thue-Morse drive (TMD) \cite{Combescure1991,Takashi2021,Hongzheng2022,Das2023,Jin2024,Tiwari2024,Chen2024,Moon2024,Tiwari2024TM},  derived from the aperiodic Thue-Morse word [Fig.~\ref{fig:01}~(a)], challenges this expectation. Reference~\cite{Nandy2017} argued that 
the TMD  has a self-similar structure and, thus, exhibits an emergent (Floquet) time-translation symmetry, precluding CHSE --- see Fig.~\ref{fig:01}~(b) for an example of seemingly nonergodic dynamics. If these predictions are correct, this represents a surprising exception to the MEP and potentially presents  new routes to stabilize driven quantum systems against the deleterious effects of heating.

In this work, we carefully analyze the dynamics of the  TMD and through it unravel a new class of quantum ergodic dynamics which we dub {\it critically slow complete Hilbert-space ergodicity} (CS-CHSE). Specifically, we rigorously prove that a qubit driven by the TMD generically displays emergent time-translation symmetries with arbitrarily good accuracy over arbitrary durations of time; however, these symmetries are finitely lived, resulting in the system achieving CHSE in the strict infinite time limit. Our finding resolves the apparent tension between the purported  nonergodicity  claimed by Ref.~\cite{Nandy2017} in the TMD and the maximally ergodic predictions of the MEP. Furthermore, our analysis uncovers that the exploration of the full Hilbert space  occurs in a scale-free manner --- precisely, the distribution of timescales that the emergent time-translation symmetries emerge for obeys a power law and has an unbounded mean, hence the term ``critically slow."

Motivated by these findings, we numerically explore other aperiodic driving sequences and find that the phenomenology of CS-CHSE holds similarly for drives derived from certain morphic words --- words derived from repeatedly applying substitution rules on basic characters~\cite{Lothaire2002}. Thus, CS-CHSE represents a new class of quantum ergodic dynamics in which full ergodicity is eventually achieved --- as predicted by the MEP ---, but over timescales astronomically larger than those natural to the system. This adds to recent studies on quantum dynamics emphasizing how there can be a large separation of timescales between the asymptotic  {\it late-time} dynamics and a transient but potentially physically more-relevant  {\it long-time} dynamics, such as two-step~\cite{Jonay2023} and slow thermalization \cite{Yao2016,Lan2018,Han2024,Balasubramanian2024,Wang2025}. 

{\it Thue-Morse drive and CHSE}.---
We begin by defining the quantum TMD. It is derived from the so-called Thue-Morse word (TMW) \cite{oeisTM}, an infinite word on two characters $\{0,1\}$ constructed by the following simple concatenation rules \cite{Lothaire2002}. Starting from the base word $W_0$\,$=$\,$0$,  form the word $W_{n+1}$ at the $(n+1)$\nobreakdash-th level by concatenating $W_n$ with its bitwise negation $\overline{W_{n}}$ (i.e., interchanging $0$ and~$1$):
    \begin{equation}
    \label{eq:TMWparts}
        W_{n+1}=W_{n} \overline{W_{n}}.
    \end{equation}
    For example, $W_1$\,$=$\,$01$ 
    yields $W_2$\,$=$\,$0110$, then $W_3$\,$=$\,$01101001$, etc. The infinite-level word $W_\infty$ is the TMW, which is known to be aperiodic, i.e.~never eventually repeats, but is not quasiperiodic \cite{Kolaifmmode1991} \footnote{We note that in previous physics literature it has been mischaracterized as such}. 

    The quantum TMD for a $d$-dimensional quantum system is realized by applying two fixed unitaries $A$\,$,$\,$B\in \operatorname{SU}(d)$ according to the characters of $W_\infty$.  Specifically, at integer time $t$\,$=$\,$n$,  apply $A$ $(B)$ if the $n$th character of $W_\infty$ is $0$ $(1)$. Thus, time evolution of an initial state $\ket{\psi(0)}$ is  $\ket{\psi(n)}$\,$=$\,$U(n)\ket{\psi(0)}$, where 
    \begin{equation}
    \label{eq:TMD}
U(n)=\mathop{\prod^\leftarrow}\limits^{n}_{j=1}\begin{cases}A&\text{if the $j$th character of $W_\infty$ is $0$} \\ B&\text{if it is $1$}\end{cases},
    \end{equation}
    is the unitary time-evolution operator.
Above, the product is  ordered from right to left. For example,  $\ket{\psi(8)}$\,$=$\,$BAABABBA\ket{\psi(0)}$.

Recent previous works have studied  the TMD, for instance  on prethermalization timescales in the high-frequency regime \cite{Takashi2021,Jin2024,Chen2024}, and in defining so-called ``time-rondeau crystals''~\cite{Moon2024}. As mentioned, a work most relevant for us is Ref.~\cite{Nandy2017}, which studied the equilibration dynamics of free-fermionic spin chains under the TMD, and argued for the emergence of nontrivial steady states.   In our context, this amounts to a violation of CHSE, which we define next. 

CHSE refers to the property of a sequence of pure states $\{\psi(t)\}_{t\in\{0,1,2,\dots \}}$ generated in dynamics, called the {\it temporal ensemble}  \cite{Mark2024}, uniformly covering the Hilbert space~\cite{Pilatowsky2023,Pilatowsky2024} \footnote{An analogous definition can apply for dynamics over continuous time as well.}. Formally, we mean that the empirical distribution of the temporal ensemble asymptotically follows    the {\it Haar ensemble}, a collection of states $\{V \dyad{\psi(0)} V^\dagger\}_V$ where $V$ is a Haar-distributed unitary on $\operatorname{SU}(d)$. This can be probed through the 
trace distance~\footnote{The trace norm is defined as $\norm{\rho}_1=\tr\sqrt{\rho \rho^\dagger}=\sum_i{\lambda_i}$, where $\lambda_i$ are the singular values of $\rho$.}
\begin{equation}
\label{eq:trace-distance}
    \Delta_T^{(k)}\coloneqq \frac{1}{2}\norm{\rho_{T}^{(k)} - \rho_{\text{Haar}}^{(k)}}_1,
\end{equation}  
of the $k$th temporal moment of the finite-time  ensemble $\rho_T^{(k)}$\,$=$\,$\frac{1}{T}\sum_{t=0}^{T-1}(U(t)\dyad{\psi}U(t)^\dagger)^{\otimes k}$ to the corresponding Haar moment $\rho_{\text{Haar}}^{(k)}$\,$=$\,$\int\dd{V}(V\dyad{\psi}V^\dagger)^{\otimes k}$, and inquiring if at large times $\lim_{T\to \infty}\Delta_T^{(k)}=0$ for all $k$~\footnote{The trace distance $\Delta_T^{(k)}$ is used because it gives the probability of distinguishing $\ket{\psi}$ from a Haar-random state using an optimal, possibly entangling measurement over $k$ copies, so the requirement $\lim_{T\to \infty}\Delta_T^{(k)}$\,$=$\,$0$ for all $k$ means that the late-time state  is indistinguishable from Haar-random, even with access to an arbitrarily large number of system replicas and entangling measurements.}. In  quantum information  parlance, one asks if the temporal ensemble forms a {\it state design}. 

Here, we will actually study the stronger notion of ergodicity of the temporal ensemble of unitaries themselves, namely ask if the statistical properties of the time-evolution operators $\{U(t)\}_t$ of the TMD are close to those of Haar random unitaries. Specifically we consider the $k$th-moment time-averaging channel $\mathcal{N}^{(k)}_T[\,\cdot\,]$\,$=$\,$\frac{1}{T}\sum_{t=0}^{T-1} U(t)^{\otimes k}[\,\cdot\,]U(t)^{\dagger \otimes k}$ and compare its similarity to the corresponding $k$th-moment Haar averaging channel $\mathcal{N}_{\mathrm{Haar}}^{(k)}[\, \cdot\,]$\,$=$\,$\int\dd V V^{\otimes k}[\,\cdot\,]V^{\dagger \otimes k}$, for example through the Hilbert-Schimdt distance 
\begin{align}
\label{eq:defintionofHSD}
D_T=\norm{\mathcal{N}^{(k)}_{T}-\mathcal{N}^{(k)}_{\mathrm{Haar}}}_2.
\end{align}
If $\lim_{T\to \infty} D_T$\,$=$\,$0$ for all $k$, then this property of the     unitary ensemble is dubbed {\it complete unitary ergodicity} (CUE)~\cite{Pilatowsky2024}. In quantum information parlance, this is asking whether the ensemble of time-evolution operators forms a {\it unitary design}. Since the $k$th moment of the finite-time state ensemble can be recovered via $\rho_T^{(k)}=\mathcal{N}^{(k)}_T[ |\psi(0)\rangle \langle \psi (0)|^{\otimes k}]$ and similarly for the Haar moment, it is clear CUE implies CHSE. 
 
\begin{figure}
    \centering
    \includegraphics[width=1\columnwidth]{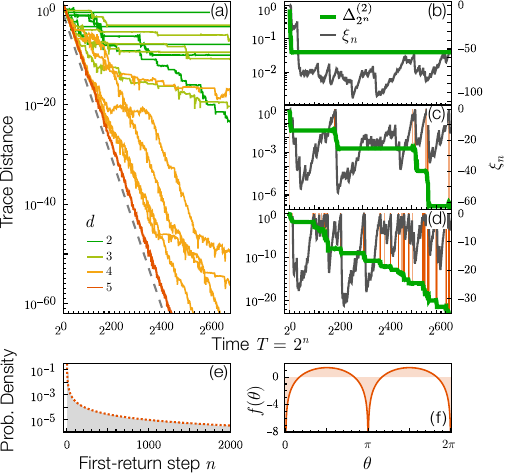}
    \caption{
    Trace distance $\Delta_{T=2^n}^{(k)}$ for $k$\,$=$\,$2$ between temporal and Haar distribution for the TMD (each line is a different realization
    of Haar-randomly sampled basic unitaries $A$\,$,$\,$B$). (a) Trace distances for various $d$-level systems. A power-law decay $T^{-1/2}$ is shown in a dashed line for reference, which is tracked for large $d$. However, plateaus appear for small $d$.  (b)-(d) Different instances of trace distance   for $d$\,$=$\,$2$ (green line) and  distance $\xi_n$ (gray line). 
    Orange vertical lines indicate times at which $\xi_n>-1.$ (e) Distribution of first-return steps of the quasirandom walk on $\xi_n$, over random initial $\theta_1$ and $\xi_1$\,$=$\,$-1/2$. 
 The orange dotted line is $\propto$\,$n^{-3/2}$.
     \label{fig:02} (f) Plot of  step-length function $f(\theta)$\,$=$\,$\log(4\sin(\theta)^2)$.}
\end{figure}

{\it Transient time-translation symmetries in the TMD}.---
First, we illustrate the basic phenomenology seen  in numerical simulations of the TMD~\cite{SM}  where the basic unitaries $A,B$\,$\in$\,$\operatorname{SU}(d)$ are chosen Haar randomly. Figure~\ref{fig:02}~(a) shows the trace distance $\Delta_T^{(k)}$ for $k$\,$=$\,$2$. For larger dimension $d$, there is a power-law behavior  $\sim$\,$1/\sqrt{T}$, which provides strong numerical evidence for CHSE. However, for small $d$, and particularly for $d$\,$=$\,$2$~[Figs.~\ref{fig:02} (b)-(d)], the decay is not smooth, and, in fact, there are surprisingly many intermediate long-lived plateaus, some up to times $2^{600}$~(!), the limit of our finite-time numerics. 
Despite the astronomically large times reached by our simulations (see~\cite{SM} for details), it is unclear if these plateaus survive indefinitely or if there is eventual decay at even later times, such that CHSE is achieved. 

{\it Dynamics of unitaries}.---We now turn to our rigorous analysis of the asymptotic behavior of the TMD.
A starting observation is that we may simplify the analysis of the TMD by concentrating on exponential times $t$\,$=$\,$2^n$, i.e., study the sequence of unitaries $A_n \coloneqq U(2^n)$. We find $A_n$ may be generated by repeated applications of a dynamical map $(A_n,B_n)$\,$=$\,$\Phi(A_{n-1},B_{n-1})$ where $B_n$ is a  partner sequence with all $A$s swapped with $B$s, which we call the `complementary TMD'. 
Specifically, define the map $\Phi\colon\operatorname{SU}(d)$\,$\times$\,$\operatorname{SU}(d)$\,$\to$\,$\operatorname{SU}(d)$\,$\times$\,$\operatorname{SU}(d)$ where $\Phi(A,B)$\,$:=$\,$(BA$\,$,$\,$AB)$; then 
\begin{equation}
\label{eq:AnBn}
    (A_n,B_n)=\Phi^n(A,B),
\end{equation} 
which is equivalent to the  recursion relations $A_{n+1}$\,$=$\,$B_nA_n$ and $B_{n+1}$\,$=$\,$A_nB_n$~\cite{Nandy2017,SM}.

Using this map, the origin of the plateauing structure of the trace distance  can be intuitively understood by analyzing the iterates $A_n$\,$,$\,$B_n$.  Suppose (for some reason) $A_n\approx B_n$ within a range $n$\,$\in$\,$[n_*,N_*]$. Then, defining  $V_*$\,$=$\,$A_{n_*}$, we have $\Phi(A_{n_*},B_{n_*})$\,$\approx$\,$(V_*^2,V_*^2)$ and subsequently, the evolution operator satisfies $U(T=2^{n})$\,$=$\,$A_n$\,$\approx$\,$V_*^{2^n}$. In other words, between times $2^{n_*}$ and $2^{N_*}$, the system behaves like a Floquet system, evolving with the repeated application of $V_*$.
This emergent discrete time-translation symmetry would halt the further exploration of the Hilbert space, due to quasienergy conservation~\cite{Pilatowsky2024}. 

This preliminary analysis suggests  that the emergent Floquet dynamics seen in numerics can be understood by showing $A_n$\,$\approx$\,$B_n$ for a large range of $n$. 
Thus, the distance $\xi_n$\,$\coloneqq$\,$\xi(A_n,B_n)$\,$<$\,$0$ between the iterates, where
\begin{equation}
\label{eq:DistanceBetweenUnitaries}
    \xi(A,B) \coloneqq  \log\left(\tfrac{1}{2}\abs{1-\tfrac{1}{d}\tr(A^\dagger B)} \right),
\end{equation}
is a key quantity to consider: it governs the accuracy of the emergent time-translation symmetries in exponential precision (we will find that it is natural to measure the distance in logarithmic scale). 
Indeed, $\xi_n$ for a qubit system is plotted in Fig.~\ref{fig:02}(b)-(d), where two salient features are observed: When $\xi_n$ is close to $0$ (vertical orange lines), the trace distance $\Delta_{2^n}^{(k)}$ appears to drop;  while when $\xi_n$ is large,  the trace distance plateaus, as expected, confirming its important role in dynamics.  However, $\xi_n$ is also seen to behave rather erratically, as if driven by a random process: Over time there are multiple large excursions from the origin but also multiple recurrences back to it. Understanding which process dominates (if any), is paramount to understanding whether CHSE is achieved. 

{\it Quasirandom walk of $\xi_n$ and CHSE}.---
We find that for a system consisting of a single qubit, the behavior of the distance $\xi_n$, which we interpret as the position of a `particle' living on the semi-infinite real line $(-\infty$\,$,$\,$0]$, can be exactly captured by a dynamical map in which the position and an angular variable $\theta_n$\,$\in$\,$[0$\,$,$\,$2\pi)$ are coupled:
\begin{align}
    \label{eq:rhothetamap}
        \xi_{n+1}&=\xi_n+\log(1-e^{\xi_n}) + f(\theta_n) \\
        \theta_{n+1}&=\arg(e^{i 2\theta_n} + (1-e^{i2\theta_n})e^{\xi_n}),\nonumber 
    \end{align}
where  $f(\theta)$\,$=$\,$\log(4\sin(\theta)^2)$. This map is derived from the map on unitaries $\Phi$~\cite{SM}. 
The initial points $\xi_1$\,$,$\,$\theta_1$ are related to the starting choice of unitaries $A$\,$,$\,$B$ whose precise expressions and values are not important  in determining the generic long-time behavior of the map. 
Alternatively, the dynamics can equivalently be represented as a nonlinear map $c_{n+1}$\,$=$\,$c_{n}^2+1-\abs{c_n}^2$ of a single complex variable in the unit disk $|c_n|\leq 1$,  which accords a complementary geometrical viewpoint of the TMD in terms of motion within the disk (see \cite{SM} for details). 
 
One may gain more physical insight into Eq.~\eqref{eq:rhothetamap} when the particle is far from the origin, so that the factor $e^{\xi_n}$ is small and can be ignored; then the angular part decouples: 
      \begin{subequations}
        \begin{align}
        \xi_{n+1}&\approx\xi_n + f(\theta_n) \label{eq:drivenwalk} \\
        \theta_{n+1}&\approx 2\theta_n \mod 2\pi. \label{eq:doubleanglemap} 
\end{align}
\end{subequations}
Equation~\eqref{eq:doubleanglemap} is the angle doubling map, which is well known to be chaotic, in the sense that the angles $\theta_n$ behave like random numbers for a generic initial condition. These random numbers, in turn, drive the position of the particle with the length of each step given by $f(\theta_n)$; thus  the iteration in Eq.~\eqref{eq:drivenwalk} describes a quasirandom walk, explaining qualitatively the behavior seen in Fig.~\ref{fig:02}~(b)-(d). 
Of course, the exact equations of motion in Eq.~\eqref{eq:rhothetamap} feature an additional `soft-wall' constraint $\log(1-e^{\xi_n})$ preventing the particle from crossing the origin.  

The equations of motion~\eqref{eq:rhothetamap} also provide a quantitative understanding of the steplike structure seen in the trace distance in Fig.~\ref{fig:02}~(b)-(d). Key is the function $f(\theta)$ [Fig.~\ref{fig:02}~(f)]: It is upper bounded by $\log(4)$, implying   a finite maximum increase in $\xi_n$ per step; but at the same time is unbounded from below, diverging to $-\infty$ at $\theta$\,$=$\,$0$\,$,$\,$\pi$, implying a potentially unbounded decrease in $\xi_n$ per step.  
Thus upon incurring a single large negative jump in $\xi_n$, the system becomes almost `Floquet-like' so that $\Delta^{(k)}_{2^n}$   plateaus,  and requires many small positive steps to conspire together to increase the value of  $\xi_n$ again so that $\Delta^{(k)}_{2^n}$ drops. One can  estimate the {\it distribution} of the lengths of these plateaus $\tau$ by analyzing the distribution of  the first-return step of $\xi_n$ under Eq.~\eqref{eq:rhothetamap} (i.e.,~the minimum $n>1$ such that $\xi_n\geq\xi_1$). We numerically find a power-law distribution $P(n)$\,$\sim$\,$n^{-3/2}$ [Fig.~\ref{fig:02}~(e)], same as for a random walk in 1D~\cite{Redner2001book}. Taking $\tau=2^n$, the distribution of plateau lengths is, thus, $P(\tau)$\,$\sim$\,$\tau^{-1} \log(\tau)^{-3/2}$, a heavy-tailed distribution with unbounded mean.

Turning now to the important question of whether the random walker wanders off to (negative) infinity or remains near its (finite) starting position at late times:  interestingly,  because the function  $f(\theta)$ governing the step size has   zero mean and finite variance, the random walk belongs to the Gaussian universality class, and so {\it both} generically happen infinitely often, rigorously captured by the following lemma. 
\begin{lemma}
\label{lemm:lem01}
   For almost any initial pair  $(\xi_1,\theta_1)$~\textnormal{\footnote{with respect to the uniform measure on the semi-infinite cylinder}}, the dynamical map  Eq.~\eqref{eq:rhothetamap} has different subsequences $\{n_m\}$ which
   \begin{enumerate}
       \item[(i)] \label{lemm:02parti} converge back to the initial point $\lim_{m\to \infty}(\xi_{n_m},\theta_{n_m}) = (\xi_1,\theta_{1})$ and 
       \item[(ii)] \label{lemm:02partii}wander off to infinity $\lim_{m\to \infty}{\xi_{n_m}}=-\infty$.
   \end{enumerate}
\end{lemma}
 \noindent The proof, taken from Ref.~\cite{NazarovMathoverflow2024}, is presented in the Supplemental Material~\cite{SM}. 
 
Statement (ii) of Lemma~\ref{lemm:lem01} formalizes the long timescales of emergent time-translation symmetry and nonergodicity. Conversely, statement~(i) guarantees that despite these long timescales, eventual uniform ergodicity is achieved: We further show that each recurrence near the initial point $(\xi_1,\theta_1)$ is accompanied by a {\it strict decrease} in the degree of dissimilarity between the ensemble of time-evolution unitaries of the TMD to Haar random ones (proof in Supplemental Material~\cite{SM}): 

 \begin{lemma}
\label{lemm:lem02}
    Let $M_n:=D_{2^n}^2 + \bar{D}_{2^n}^2$ where $\bar{D}_{T}$ is the distance measure  Eq.~\eqref{eq:defintionofHSD} but for the complementary TMD (i.e., with $A$ interchanged with $B$).
    Then, $M_n$ is a monotone, and   strictly decreases by a constant multiplicative factor every time $(\xi_n,\theta_n)$ is sufficiently close to $(\xi_1,\theta_1)$, for almost all choices of starting unitaries~$(A,B)$. 
\end{lemma}

 Since Lemma~\ref{lemm:lem01}~(i) implies these strict decreases happen infinitely often,  $\lim_{n \to \infty} M_n = 0$ and hence  $\lim_{T \to \infty} D_T = 0$ \cite{SM}, yielding the expected result from the MEP that the TMD ultimately achieves maximal ergodicity. 
 \label{sec:TMISCUE}
\begin{theorem}
\label{theorem:CUEinTMD}
The TMD for a qubit system exhibits CUE and, thus, CHSE for almost any choice of $A,B\in \operatorname{SU}(2)$ (with respect to the Haar measure). 
\end{theorem}
We note that the technical reason  Ref.~\cite{Nandy2017} did not identify the CHSE of the TMD is their claim that $\xi_n\to-\infty$ ($\phi_m\to 0$ in their notation), which  does not hold generically [statement~(i) of Lemma~\ref{lemm:lem01}].

{\it Discussion and outlook}.---Our study of the aperiodic TMD has confirmed the applicability of the MEP: it correctly predicts that the drive, which lacks any explicit symmetries, ultimately produces an ergodic temporal ensemble of states,  uniformly distributed over the  Hilbert space. At the same time, our work has also explained how this maximal ergodicity is attained  dramatically slowly and in a scale-free manner (at least for a qubit), a dynamical phenomenon we dub CS-CHSE. Our work has direct implications for experiments probing CHSE (such as in a nitrogen-vacancy center setup, recently performed~\cite{Liu2025}), in correctly interpreting the apparent nonergodic behavior they would find at short times if they were to implement the TMD.  We leave as an open question why  the long periods of nonergodicity seem to vanish with increasing system size [Fig.~\ref{fig:01}~(a)], for which  it would be useful to understand the invariant measure of the map $\Phi$ in arbitrary dimension. 

\begin{figure}[t]
    \centering
    \includegraphics[width=1\linewidth]{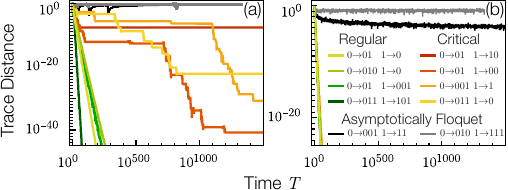}
    \caption{Trace distance $\Delta_T^{(k)}$ between the Haar distribution and temporal ensemble generated by   quantum morphic drives associated with substitution rules labeled in panel (b), for Haar-randomly chosen starting unitaries $A$\,$,$\,$B$ (same for all drives). (a) $d$\,$=$\,$2$\,$,$\,$k$\,$=$\,$2$, (b) $d$\,$=$\,$5$\,$,$\,$k$\,$=$\,$1$.}
    \label{fig:03}
\end{figure}

Beyond the TMD, it is natural to inquire about the generality of CS-CHSE. To that end, we perform a preliminary numerical study of aperiodic discrete quantum drives constructed by applying basic unitaries $A$~$(B)$ according to the characters $0$~$(1)$ appearing in a binary morphic word --- an infinite word generated by iterating substitution rules $0$\,$\to$\,$w_0$\,$,$\,$1$\,$\to$\,$w_1$ starting from $0$, where $w_0$\,$,$\,$w_1$ are some fixed subwords~\cite{Lothaire2002}. For example, the TMW studied in this work is the morphic word defined by $0$\,$\to$\,$01$, $1$\,$\to$\,$10$. 

In Fig.~\ref{fig:03} we plot the trace distances between the temporal and Haar ensembles for some morphic drives constructed from words with choices of $w_0$ and $w_1$ of length at most 3. We find three distinct behaviors: first, drives which display ``regular" CHSE~(green lines), in the form of a power-law decay, regardless of dimension; these include the generalized Fibonacci drives ($0$\,$\to$\,$0^m1$, $1$\,$\to$\,$0$) studied in \cite{Pilatowsky2023} and  generalizations like  the Octonacci drive $(0$\,$\to$\,$010$, $1$\,$\to$\,$0)$~\cite{oeisOcto}. Second, drives which appear to display CS-CHSE (yellow-red lines) for a qubit~[Fig.~\ref{fig:03}(a)], which include the TMD, but also drives from other sequences, like the period-doubling word $(0$\,$\to$\,$01, 1$\,$\to$\,$00)$~\cite{oeisPeriodDoubling}. This shows that CS-CHSE is not particular to the TMD. While in the present examples CS-CHSE is seen to vanish with increasing Hilbert-space dimension~[Fig.\ref{fig:03}~(b)], understanding if there are more general morphic drives wherein CS-CHSE survives in the many-body regime would be an important direction for future work. Third, there is a small proportion of drives (black and gray lines) which exhibit a yet slower decay of the trace distance, present for any dimension. However, these have arguably a trivial origin, as the underlying words have arbitrarily long streaks of repeated characters (e.g.~for $0$\,$\to$\,$001, 1$\,$\to$\,$11$, each substitution doubles the length of the blocks of 1's)~\footnote{This is unlike the TMW, which is {\it cubefree}~\cite{oeisTM} ---no word appears repeated more that twice.}; hence, we term  them `asymptotically Floquet drives'. Nevertheless, given that asymptotically Floquet drives are strictly speaking aperiodic, it is an open and nontrivial question whether  CHSE is ultimately achieved or not in their dynamics. We see that there is a rich interplay between the combinatorial structure and complexity  of the aperiodic morphic words  and the degree of ergodicity in the quantum drives they generate, which deserves to be better studied.   

\begin{acknowledgments}
{\it Acknowledgments}.---We are immensely grateful to Fedor Nazarov of Kent State University for his invaluable help in the proof of Theorem~1.  We also thank S.~Nandy for insightful conversations. S.~P.~C. and S.~C. acknowledge support from the Center for Ultracold Atoms (an
NSF Physics Frontiers Center; PHY-2317134), NSF CAREER (DMR-2237244), and the Heising-Simons Foundation (grant \#2024-4851). W.~W.~H.~is supported by the National Research Foundation (NRF), Singapore through the NRF Felllowship NRF-NRFF15-2023-0008 and through the National Quantum Office, hosted in A*STAR, under its Centre for Quantum Technologies Funding Initiative (S24Q2d0009).
\end{acknowledgments}

\bibliography{references}

\end{document}

\title{
Supplemental Material: \\
Critically Slow Hilbert-Space Ergodicity in Quantum Morphic Drives
}
\author{Sa\'ul Pilatowsky-Cameo}
\affiliation{Center for Theoretical Physics, Massachusetts Institute of Technology, Cambridge, Massachusetts 02139, USA}
\author{Soonwon Choi}
\affiliation{Center for Theoretical Physics, Massachusetts Institute of Technology, Cambridge, Massachusetts 02139, USA}
\author{Wen Wei Ho}
\affiliation{Department of Physics, National University of Singapore, Singapore 117551}
\affiliation{Centre for Quantum Technologies, National University of Singapore, 3 Science Drive 2, Singapore 117543}
\maketitle

    In this supplemental material, we present the technical calculations and proofs which support our results and enable our numerical calculations. In Sec.~\ref{sec:recursiveeqs} we show that the time-averaging channels of the Thue-Morse drive (TMD) and general quantum morphic drives follow a recursive structure which enables our late-time numerical simulations. In Sec.~\ref{app:disk_map}, we show the transformation that allows to map the Thue-Morse map in $\operatorname{SU}(2)$ to a two-dimensional dynamical map. In Sec.~\ref{sec:recurrenceofdiskmap} we present the proof from Ref.~\cite{NazarovMathoverflow2024} of Lemma~1 of the main text. Finally, Sec.~\ref{sec:prooftheorem} contains the full proof of Theorem~1 of the main text, the complete unitary ergodicity (CUE) of the single-qubit TMD.

    \section{Recursive equations for the time-averaging channel }
We explain that the time-averaging channels 
$\mathcal{N}_T[\,\cdot\,]=\frac{1}{T}\sum_{t=0}^{T-1} U(t)^{\otimes k}[\,\cdot\,]U(t)^{\dagger \otimes k}$ for the TMD and general quantum morphic drives follow a recursive relation.  We then show how we can leverage this recursive structure to compute the temporal moments for times which are exponential in the number of computational operations.
\label{sec:recursiveeqs}
\subsection{Thue-Morse drive}
We begin with the relations for time-averaging channel of the TMD. Recall that the evolution operator at exponential times  $A_n=U(t=2^n)$ follows the recursion $A_{n+1}=B_nA_n$, $B_{n+1}=A_nB_n$. 
\begin{prop}
\label{prop:time-av-channel}
    For any moment $k\in \mathbb{N}$, the time-averaging channel $\mathcal{N}_{2^n}^{(k)}$ 
    for the TMD satisfies the following recursive equations
    \begin{align}
        \label{eq:recursiveeqstime}\mathcal{N}^{(k)}_{2^{n+1}}&=\tfrac{1}{2}(\mathcal{N}^{(k)}_{2^n}+\overline{\mathcal{N}}^{(k)}_{2^n}\circ \mathcal{A}_n) \\ \overline{\mathcal{N}}^{(k)}_{2^{n+1}}&=\tfrac{1}{2}(\overline{\mathcal{N}}^{(k)}_{2^n}+\mathcal{N}^{(k)}_{2^n}\circ \mathcal{B}_n), \nonumber
    \end{align}
    where  $\mathcal{A}_n[\,\cdot\,]=A_n^{\otimes k}[\,\cdot\,]A_n^{\dagger \otimes k}$ and $\mathcal{B}_n [\,\cdot\,]=B_n^{\otimes k}[\,\cdot\,]B_n^{\dagger \otimes k} $, and initial conditions $        \mathcal{N}^{(k)}_{2^0}=\mathcal{A}_0$ 
     and $\overline{\mathcal{N}}^{(k)}_{2^0}=\mathcal{B}_0$. 
\end{prop} \noindent 
\begin{proof}
    
The result follows simply from the recursive construction of the TM word \begin{equation}
\label{eq:TMWparts}
    W_{n+1}=W_{n} \overline{W_{n}},
\end{equation} where recall $\overline{W_{n}}$ is the bitwise negation of the word $W_{n}$. To make this explicit, we will introduce some notation. Let $\omega_j$ be the $j$th character of the TM word $W_\infty$. Note that this is also the $j$th character of $W_n$ as long as $j\leq 2^n$. Let $\overline{\omega}_j$ be the bit negation of $\omega_j$. From Eq.~\eqref{eq:TMWparts}, we can see that for any $t\in[0:2^{n})\coloneqq \{0,1,\dots, 2^{n}-1\}$ we have 
\begin{align}
W_n\overline{\omega}_1\overline{\omega}_2\cdots \overline{\omega}_t&=\omega_1\omega_2\cdots \omega_{2^n+t} \\ \overline{W}_n{\omega}_1{\omega}_2\cdots {\omega}_t&=\overline\omega_1\overline\omega_2\cdots \overline\omega_{2^n+t},\nonumber
\end{align}
and, consequently,
\begin{align}
\label{eq:UandUbar}
    \overline{U}(t)A_n &= U(2^n+t)
    \\{U}(t)B_n &= \overline{U}(2^n+t),\nonumber
\end{align}
where $\overline{U}(t)$ is defined the same way as $U(t)$, but interchanging $A$ for $B$.
Let us denote by \begin{equation}
\label{eq:partialaveraging}
    \mathcal{N}_{[T_i:T_f)}[\, \cdot\,]=\frac{1}{T_f-T_i}\sum_{t=T_i}^{T_f-1} U(t)^{\otimes k}[\, \cdot\,] U(t)^{\dagger \otimes k},
\end{equation} the time-averaging channel between times $T_i$ and $T_f-1$,
and similarly $\overline{\mathcal{N}}_{[T_i:T_f)}$ with $\overline{U}(t)$ instead of $U(t)$. Note that  $\mathcal{N}_{T_f}=\mathcal{N}_{[0:T_f)}$. Then from Eqs.~\eqref{eq:UandUbar} we see that
\begin{align}
\label{eq:NandNbarpartialtime}
\overline{\mathcal{N}}_{[0:2^{n})} \circ \mathcal{A}_n&=\mathcal{N}_{[2^n:2^{n+1})}\\
{\mathcal{N}}_{[0:2^{n})} \circ \mathcal{B}_n&=\overline{\mathcal{N}}_{[2^n:2^{n+1})}.\nonumber
\end{align}
We obtain Eqs.~\eqref{eq:recursiveeqstime} from Eqs.~\eqref{eq:NandNbarpartialtime} by noting that \begin{align*}
\mathcal{N}_{2^{n+1}}&=\tfrac{1}{2}\big(\mathcal{N}_{[0:2^{n})} + \mathcal{N}_{[2^n:2^{n+1})}\big)\\\overline{\mathcal{N}}_{2^{n+1}}&=\tfrac{1}{2}\big(\overline{\mathcal{N}}_{[0:2^{n})} + \overline{\mathcal{N}}_{[2^n:2^{n+1})}\big).\qedhere\end{align*}

\end{proof} 
\subsection{Generalization to other morphic words}
We generalize the recursive equations for the time-averaging channel of the TMD to drives generated by other morphic words, which we call {\it quantum morphic drives}. We consider a general substitution rule $0\mapsto \sigma(0)$, $1\mapsto \sigma(1)$ on a binary alphabet\footnote{We could also consider substitution in larger alphabets: the generalization is straightforward, with more driving unitaries.}, where $\sigma(1)$ and $\sigma(0)$ are finite binary words. To generate an infinite morphic word, we repeatedly apply the substitution starting on $0$,  as we illustrate in Table~\ref{tab:subs} for the Thue-Morse, Fibonacci, and period-doubling words. In general, this iteration procedure has a unique well-defined limiting word $W=\lim_{n\to\ \infty} \sigma^n(0)$ as long as $\sigma(0)$ starts with $0$ \cite{Lothaire2002}.

\begin{table}[]
    \centering
   \begin{tabular}{|c|ll|clclclclcll|}
    \hline
    Name & \multicolumn{2}{c|}{Substitution} & \multicolumn{11}{c|}{Iteration} \\
    \hline
    Thue-Morse \cite{oeisTM} & $\sigma(0)=01$ & $\sigma(1)=10$ & $0$&$\mapsto$&$01$ & $\mapsto$ & $0110$ & $\mapsto$ & $01101001$&$\mapsto$&$0110100110010110$&$\mapsto$&$\dots$ \\
    \hline
    Fibonacci \cite{oeisFibb} & $\sigma(0)=01$ & $\sigma(1)=0$ & $0$&$\mapsto$&$01$ & $\mapsto$ & $010$ & $\mapsto$ & $01001$ & $\mapsto$ & $01001010$ & $\mapsto$ & $\dots$ \\
    \hline
    Period-Doubling \cite{oeisPeriodDoubling}& $\sigma(0)=01$ & $\sigma(1)=00$ & $0$&$\mapsto$&$01$ & $\mapsto$ & $0100$ & $\mapsto$ & $01000101$ & $\mapsto$ & $0100010101000100$ & $\mapsto$ & $\dots$ \\
    \hline
\end{tabular}

    \caption{Substitution rules and first few of their iterations starting on $0$ for the Thue-Morse, Fibonacci and period-doubling words. }
    \label{tab:subs}
\end{table}

    We generate quantum dynamics by following the characters of $W$ in time. At time $t=n$, we apply the unitary $V^{[0]}=A$ if the $n$'th character is $0$ and the unitary $V^{[1]}=B$ if the $n$'th character of $W$ is $1$, i.e., the evolution operator is given by
    \begin{equation}
        U(t)=\mathop{\prod^\leftarrow}\limits^{t}_{j=1}V^{[W_j]},
    \end{equation}
    where $W_j\in\{0,1\}$ is the $j$th character of the morphic word $W$. To probe CHSE, we care about the temporal moments, which can be obtained  through the time-averaging channel $\mathcal{N}^{[0]}_T[\,\cdot\,]=\frac{1}{T}\sum_{t=0}^{T-1} U(t)^{\otimes k}[\,\cdot\,]U(t)^{\dagger \otimes k}$. We use the superscript $^{[0]}$ to differentiate this channel from the complementary channel (obtained by exchanging $A$ and $B$) which we denoted as $\overline{\mathcal{N}}_T$ for the TMD, but here denote as $\mathcal{N}^{[1]}_T$ to facilitate writing equations below. We consider these channels at the {\it word times}, $T=S_n^{[b]}$, equal to the length of $\sigma^n(b)$, ($b\in\{0,1\}$). These can be computed from the equations
\begin{equation}
    S_{n+1}^{[b]}=\sum_{j=1}^{\ell(b)} S^{{[\sigma(b)_j]}}_n, \quad S^{[b]}_0=1,
\end{equation}
where $\ell(b)$ denotes the length of the word $\sigma(b)$ and $\sigma(b)_j$ denotes its $j$th character. For example, for the Fibonacci substitution, $S^{[b]}_{n}$ are just the Fibonacci numbers (with an index offset depending on $b$), and for the Thue-Morse substitution, $S_n^{[b]}=2^n$. The numbers $S_n^{[b]}$ grow exponentially with the index $n$ for nontrivial substitutions.

We now consider the sequence of unitary evolution operators at the word times $V^{[0]}_n=U(T=S_n)$ (for the TMD we denoted this as $A_n$). These follow a recursive equation, together with a partner sequence $V^{[1]}_{n}=B_n$, 
\begin{equation}
   V^{[b]}_{n+1}=V_n^{[\sigma(b)_{\ell(b)}]}\cdots V_n^{[\sigma(b)_{2}]} V_n^{[\sigma(b)_{1}]},\quad V_0^{[b]}=V^{[b]},
\end{equation}
which follows readily from the substitution rule. Importantly, the 
 time-averaging channels also obey a  recursion relation
\begin{equation}
    \label{eq:generalmorphicchannelrelation}\mathcal{N}^{[b]}_{S_{n+1}}=\sum_{j=1}^{\ell(b)} \frac{S_{n}^{[\sigma(b)_j]}}{S_{n+1}^{[b]}} \mathcal{N}_{S_{n}}^{[\sigma(b)_j]}\circ \mathcal{V}_n^{[\sigma(b)_{j-1}]}\circ \cdots \circ\mathcal{V}_n^{[\sigma(b)_{1}]},
\end{equation}
with the unitary channels $\mathcal{V}_n^{[b]}[\,\cdot\,]={\big(V^{[b]}_n\big)^{\otimes k}}[\,\cdot\,]{\big(V^{[b]}_n\big)^{\dagger \otimes k}}$.
Equations~\eqref{eq:recursiveeqstime} are the particular case of Eq.~\eqref{eq:generalmorphicchannelrelation} for the Thue-Morse substitution. For the Fibonacci substitution, Eq.~\eqref{eq:generalmorphicchannelrelation} reduces to Eq. (7) of Ref.~\cite{Pilatowsky2023}.

\subsection{Numerical simulations}
Equations~\eqref{eq:recursiveeqstime} and the more general Eq.~\eqref{eq:generalmorphicchannelrelation} allow to calculate time averages for times which are exponential in the number of computational operations. We compute the channels $\mathcal{N}_{2^{n}}$ (or $\mathcal{N}_{S_n}^{[0]}$) as matrix operators in vectorized space and then apply them to the initial state $\dyad{\psi}^{\otimes k}$ to obtain the finite-time average $\rho_{T}^{(k)}=\mathcal{N}_{T}[\dyad{\psi}^{\otimes k}]$. The Haar average can be can be computed analytically~\cite{Harrow2013},
\begin{align}
\rho_\mathrm{Haar}^{(k)}=\frac{\sum_{\pi\in\mathcal{S}_k} \sum_{\alpha_1 \cdots \alpha_k}\dyad{\alpha_1 \alpha_2 \cdots \alpha_k}{\alpha_{\pi(1)}\cdots\alpha_{\pi(k)}}}{d(d+1)\cdots (d+k-1)},\label{eq:haarensemble}
\end{align} with $\ket{\alpha}$  any orthonormal states labeled by $\alpha\in\{0,\dots,d-1\}$, and  $\mathcal{S}_k$  the symmetric group $k$ elements, containing the permutations $\pi\colon\{1,2,\dots,k\} \to \{1,2,\dots,k\}$. Thus, we can readily access the trace distance between the temporal and Haar moments at times $S_n$ ($2^n$ for the TMD). 

Because of the recursive structure, numerical error is amplified in every step. 
We overcome this by using high-precision numbers.
\section{Quasirandom walk on $\xi_n$ and  disk map}
\label{app:disk_map}
We begin by explaining how the dynamics of the map $\Phi:\operatorname{SU}(2)\times \operatorname{SU}(2)\to \operatorname{SU}(2)\times \operatorname{SU}(2)$ given by $\Phi(A,B)=(BA,AB)$ can be effectively reduced to the dynamics of the disk map $F:\mathbb{D}\to \mathbb{D}$ given by $F(c)=c^2+1-|c|^2$, where $c$ is a complex number in the unit disk $\mathbb{D}\subset \mathbb{C}$. Consider the initial pair of unitaries $(A,B)$. 
 Then we may decompose $$A_n=\exp\big(-i \alpha_n \hat{a}_n\cdot \bar{\sigma}\big),\quad\quad B_n=\exp\big(-i \beta_n \hat{b}_n\cdot \bar{\sigma}\big)$$ with $\hat{a}_n,\hat{b}_n$ normalized vectors in $\mathbb{R}^3$, $\alpha_n \in [0,\pi)$, and $\bar{\sigma}$ the Pauli vector. For $n\geq 1$, which we henceforth assume, the unitaries have the same rotation angle $\alpha_n=\beta_n$ because $\tr(A_n)=\tr(B_{n-1}A_{n-1})=\tr(A_{n-1}B_{n-1})=\tr(B_n)$. 
 
 From $(A_{n+1},B_{n+1})=(B_nA_n,A_nB_n)$ we obtain the following recursive equations \cite{Nandy2017}
\begin{align}
\label{eq:recequations}
\cos \alpha_{n+1}&= \cos(\alpha_n)^2-\sin(\alpha_n)^2  \hat{a}_n \cdot \hat{b}_n,\\ \sin\alpha_{n+1} \hat{a}_{n+1}&=  \sin \alpha_n \cos \alpha_n( \hat{a}_n+\hat{b}_n) \nonumber
 +\sin(\alpha_n)^2  \hat{a}_n \times \hat{b}_n, \\
\sin \alpha_{n+1} \hat{b}_{n+1}&=  \sin \alpha_n \cos \alpha_n( \hat{a}_n+\hat{b}_n) -\sin(\alpha_n)^2  \hat{a}_n \times \hat{b}_n .\nonumber
\end{align}
Note that the direction of $\hat{a}_{n}+\hat{b}_{n}$ is conserved. By otherwise rotating $A,B$, we may assume that $\hat{a}_n+\hat{b}_n$ is parallel to $\hat{z}$, meaning that $\hat{b}_n=(-a_n^{X},-a_n^{Y},a_n^{Z})$. Then, by adding and subtracting the second and third lines in Eqs.~\eqref{eq:recequations} we obtain
\begin{align}
    \sin(\alpha_{n+1})(a^X_{n+1},a^Y_{n+1})&=2a_n^Z\sin(\alpha_n)^2(a_n^Y,-a_n^X)\label{eq:xypart}\\
    \label{eq:azpart}\sin(\alpha_{n+1})a^Z_{n+1}&=2a_n^Z\sin(\alpha_n)\cos(\alpha_n)\\
    \cos(\alpha_{n+1})&=1-2(\sin(\alpha_n)a_n^{Z})^2. \label{eq:alphapart}
\end{align}
The $a^X$ and $a^Y$ components  have dynamics in Eq.~\eqref{eq:xypart} which do not affect the $a^Z$ component or $\alpha$, so we focus $a^Z$ and $\alpha$ only. Eqs.~\eqref{eq:azpart} and \eqref{eq:alphapart} can be written in terms of the complex variable $c_n=\cos(\alpha_n)+i\sin(\alpha_n)a_n^Z$, as $ c_{n+1}=c_n^2+1-|c_n|^2$. 

\begin{figure}
    \centering
    \includegraphics[width=0.5\linewidth]{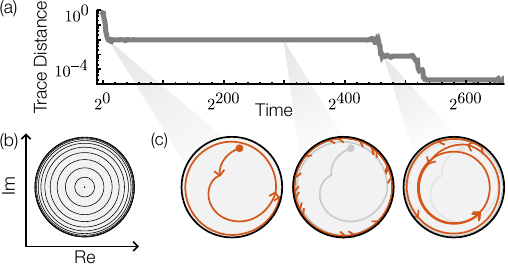}
    \caption{(a)~Plot of the trace distance between the temporal distribution and the  uniform distribution $\Delta_{T=2^n}$.     (b)~Invariant measure of the disk map, represented by concentric circles equally spaced in area. (c) Disk map: when the dynamics of the disk map are close to the boundary, the trace distance plateaus, due to the emergence of a transient time-translation symmetry, which breaks when the dynamics come back closer to the center of the disk. }
    \label{fig:SM1}
\end{figure}

The behavior of the disk map is  illustrated in Fig.~\ref{fig:SM1}~(c). When the disk coordinate is near the boundary $\abs{c_n}\approx1$, we have $A_n\approx B_n$, which results in an emergent time-translation symmetry. The invariant measure of the disk map [derived below in Sec~\ref{sec:recurrenceofdiskmap} and depicted in Fig.~\ref{fig:SM1}~(b)] is heavily concentrated near the boundary, which results in long periods of time during which the drive approximates a Floquet system. We note that Ref.~\cite{Nandy2017} claimed that $\phi_n\coloneqq\mathrm{arccos}(\hat{a}_n\cdot \hat{b}_n)\to 0$, or equivalently $\lim_{n\to \infty} |c_n|^2=1$. However, we prove in Lemma~\ref{lemm:lem01} below that this limit generically does not exist: the point in the disk repeatedly comes back arbitrarily close to the starting point, so the Floquet approximation eventually breaks.

The disk map can be equivalently written in terms of $\xi_n=\log(1-|c_n|^2)$ and $\theta_n=\arg(c_n)$, producing
    \begin{align}
    \label{eq:rhothetamap}
        \xi_{n+1}&=\xi_n+\log(1-e^{\xi_n}) + f(\theta_n),  \\
        \theta_{n+1}&=\arg(e^{i 2\theta_n} + (1-e^{i2\theta_n})e^{\xi_n}),
    \end{align}
which are the dynamics analyzed in the main text. Note that $\abs{c_n}\approx 1$ occurs when $-\xi_n\gg 1$.

\section{Recurrence and divergence of the disk map and quasirandom walk}
\label{sec:recurrenceofdiskmap}
Here, we present the proof of Lemma~1 of the main text (restated below). The proof is taken entirely from Ref.~\cite{NazarovMathoverflow2024}, with only some notational changes.

\begin{lemma}
\label{lemm:lem01}[Restatement of Lemma~1]
   For almost any initial pair  $(\xi_1,\theta_1)$~(with respect to the uniform measure on the semi-infinite cylinder), the dynamical map given by Eq.~\eqref{eq:rhothetamap} has different subsequences $\{n_m\}$ which
   \begin{enumerate}
       \item[(i)] converge back to the initial point $\lim_{m\to \infty}{(\xi_{n_m},\theta_{n_m})}=(\xi_1,\theta_{1})$, 
       \item[(ii)] wander off to infinity $\lim_{m\to \infty}{\xi_{n_m}}=-\infty$.
   \end{enumerate}
\end{lemma}

Lemma~\ref{lemm:lem01} can be equivalently phrased in terms of the disk map (see Sec.~\ref{app:disk_map})
\begin{equation}
    F(c)=c^2+1-|c|^2, \label{eq:diskmapiterates}
\end{equation} via the transformation $c_n=\sqrt{1-e^{\xi_n}}e^{i\theta_n}$. We show that for almost any initial point in the disk $c_1\in\mathbb{D}$ (with respect to the usual measure in the disk\footnote{Note that zero-measure sets in the disk correspond to zero-measure sets in the cylinder ($\xi$,$\theta$).}), there exist different subsequences such that
   \begin{enumerate}
       \item[(i)] converge back to the initial point $\lim_{m\to \infty}{c_{n_m}}=c_1$, 
       \item[(ii)] converge to the boundary of the disk  $\lim_{m\to \infty}{|c_{n_m}|}=1$.
   \end{enumerate}

The first step is to identify a measure $\mu(c)$ on the disk which is preserved by the transformation $F$, meaning that $\mu(F^{-1}(E))=\mu(E)$ for any measurable $E\subseteq \mathbb{D}$. The transformation $F(x+iy)=1-2y^2 + i(2xy)$ has two right-inverse functions \begin{equation}
         F^{-1}_\pm(x+iy)=\pm\left(\frac{y}{\sqrt{2(1-x)}}+i\sqrt{\frac{1-x}{2}}\right). \label{eq:Finverses}
         \end{equation} Thus, in order to preserve the measure we require $2\mu(F(c))=\mu(c)J(c)$ where $J(c)=\det(\begin{matrix}\partial_{x} F_x  & \partial_{x} F_y \\ \partial_{y} F_x
          & \partial_{y} F_y \end{matrix})$ is the Jacobian determinant of $F$, and $c=x+iy$. We can easily compute $J(c)=8y^2=2\frac{1-|F(c)|^2}{1-|c|^2}$ from which we see that the invariant measure is $\mu(c)=\frac{\dd^2{c}}{1-|c|^2}=\frac{\dd^2{c}}{1-|c|^2}$. 
          
          We plot this measure in Fig.~\ref{fig:SM1}~(b). It is similar to that of the Poincar\'e disk, only differing by a square in the denominator. It shares the feature that the disk has an infinite measure $\mu(\mathbb{D})=\infty$.      Armed with an invariant measure, we can prove the two parts of the lemma.
\begin{proof}[Proof of part (ii) of Lemma~\ref{lemm:lem01}]
     Assume that there is a positive measure set $E$ of initial conditions that stay $\delta$-away from the boundary $(|c_n|<1-\delta)$. Define $G=\bigcap_{n\geq 1}G_n$, with $G_n=\bigcup_{m\geq n} F^m(E)$. As $F^{-1}G_n\supseteq G_{n-1}$, $G_1\supseteq E$ and $F$ is $\mu$-preserving, $\mu(G_n)\geq \mu(E)$ for all $n$. Note that $G\subseteq F^{-1} (G)$ and, so, because $F$ is invariant, $G=F^{-1} (G)$ up to some zero measure correction. Consequently $G=\bigcap_{n\geq 1} F^{-n} (G)$ up to a zero measure correction. But, because $\mu(G)>\mu(E)>0$, we have $\bigcap_{n\geq 1} F^{-n} (G)\neq \emptyset$. So we have found an initial condition $c\in G$ such that for every $n$ the preimages $F^{-n}(\{c\})$ stay $\delta$-away from the boundary. We will argue that this is impossible: it is always possible to construct a sequence of preimages that goes back to the boundary. 
     Using the inverse functions of Eq.~\eqref{eq:Finverses}, we construct a sequence $c_{-(n+1)}=F^{-1}_\pm(c_{-n})$ (backwards in time), where the signs $\pm$ are chosen as follows. At every step, chose the sign of $y_{-(n+1)}$ opposite to that of $y_{-n}$. Then $x_{-(n+1)}=\frac{y_{-n}}{2y_{-(n+1)}}<0$, so $1-\abs{c_{-n}}^2=\frac{1-\abs{c_{-(n+1)}}^2}{2(1-x_{-(n+1)})}\leq \frac{1-\abs{c_{-(n+1)}}^2}{2}$. Consequently, $\lim_{n\to \infty} |c_{-n}| =1 $.
\end{proof}
\begin{proof}[Proof of part (i) of Lemma~\ref{lemm:lem01}]
       We shall show that for every Borel set $E\subset\mathbb D$, one has
\begin{equation}
    \mu\Big(E\setminus \bigcup_{j\ge 1} F^{-j}(E)\Big)=0. 
\end{equation}
This implies that almost any point in $E$ returns to $E$ infinitely often, which implies the existence of subsequence that converges to the initial point (taking $E$ to an arbitrarily small neighborhood around the initial condition).
 To that aim, suppose that $E\setminus \bigcup_{j\ge 1} F^{-j}(E)$ contains a set $G$ of positive measure that is contained in the disk $\{R(c)\le \Delta\}$ where $R(c)=\frac 1{1-|c|^2}$. Then the sets $F^{-j}(G)$ are pairwise disjoint and have the same $\mu$-measure $\mu(G)$ each. Consider the two right inverses $F^{-1}_\pm$ given by Eq.~\eqref{eq:Finverses}. We have $R(F^{-1}_{\pm}(c))=2(1-x)R(c)$ and $\mu(F^{-1}_{\pm}(E))=\frac 12\mu(E)$.

{\bf Claim 1. }{\it  Consider the random process in the disk, $z_0=z$, $z_{k}=F^{-1}_{\pm}(z_{k-1})$ where $\pm$ are chosen independently with probability $\frac 12$ each. Then for every $\varepsilon>0$, there exists $N(\varepsilon)$ such that for all $n>N(\varepsilon)$, one has
\begin{equation}
    \mathbb{P}[R(z_n)>e^{\varepsilon n}R(z)]<\frac 12\,.
\end{equation}}
From Claim 1 we have $\mu(F^{-j}(G)\cap \{R<e^{\varepsilon n}\Delta\})\ge \frac{\mu(G)}2$ for $j=N(\varepsilon),\dots,n$, so we can squeeze $n-N(\varepsilon)$ disjoint sets of measure $\frac{\mu(G)}2$ into the set $\{R<e^{\varepsilon n}\Delta\}$ of measure $\pi[\varepsilon n+\log\Delta]$. If $\varepsilon<\frac{\mu(G)}{2\pi}$, that is absurd for large $n$.

The rest of the proof is devoted to proving Claim 1 above. Henceforth, we fix two constants $\delta\ll C\ll\varepsilon$, a large integer $n$ and $m\approx Cn$. We shall say that our random process is {\it out of control at step $k$} if
\begin{equation}
    R(z_k)<4^{m}e^{Ck}R(z)\,.
\end{equation}
Note that if the last out-of-control position $k$ is above $n-m$, then
\begin{equation}
R(z_n)\le 4^{n-k}R(z_k)\le 4^{2m}e^{Cn}R(z)\ll e^{\varepsilon n}R(z)\,.
\end{equation}
So, it suffices to estimate the probability of the event that the last out-of-control position is $k$ for all $k=0,\dots,n-m$ by something less than $\frac 1{2n}$. Note also that $z_0,\dots,z_m$ are all out of control ($R(z_k)$ grows at most $4$ times at each step).

Fix $k\in[m,n-m]$ now. Notice that the $k$ is an out-of-control position, but $k+1,\dots,k+m$ are not. That implies that $R(z_k)\le 4^{m}e^{Ck}R(z)$ but $R(z_j)\ge 4^m$ for $j=k,\dots,k+m$. Thus, it will suffice to show that for each $w\in\mathbb D$ with $R(w)\le 4^{m}e^{Ck}R(z)$, the conditional probability
\begin{equation}
    \mathbb{P}[k+1,\dots,k+m\text{ are all in control}\,|\, z_k=w]\le \frac 1{2n}\,. \label{eq:condprob}
\end{equation}

Let $V=\frac{w}{|w|}$. Let also $V_j$ $(j=0,\dots,m)$ be the square-root random process starting with $V_0=V$ (at every step we choose one of the $2$ square roots with probability $\frac 12$ each). The next claim is that every fully controlled path $z_k,\dots,z_{k+m}$ corresponds to its own unique path in the square root process so that $|z_{k+j}-V_j|\le 2\cdot 4^{-m}$, say. Clearly, $z_k=w$ is that close to $V_0=V$. Suppose that we have managed such coupling up to step $j$. Note that $z_k=F(z_{k+j+1})=z_{k+j+1}^2+(1-|z_{k+j+1}|^2)$ differs from $z_{k+j+1}^2$ by at most $4^{-m}$ and, thereby, $|z_{k+j+1}^2-V_j|\le 3\cdot 4^{-m}$. Now we need an elementary lemma in the unit disk. If $|a|=1$, $|b|<1$ and $|a^2-b^2|<\gamma<\frac 14$, then $|a-b||a+b|<\gamma$, whence $|a-b|<\sqrt{\gamma}<\frac 12$ or $|a+b|<\sqrt{\gamma}<\frac 12$. In the first case $|a+b|\ge\frac 32$, so $|a-b|\le \frac 23\gamma$. In the second case $|a+b|\le \frac 23\gamma$. Applying it to our situation, we see that $z_{k+j+1}$ is $2\cdot 4^{-m}$ close to one of the two square roots of $V_j$ and if we can make both choices $z_{k+j+1}=F^{-1}_{\pm}(z_{k+j})$, they are close to opposite roots of $V_j$ being the opposite numbers themselves. Thus we can couple up to $j+1$ as well.

Now, let $X(\zeta)$ be the $x$-coordinate of $\zeta$. Consider the path $V_j$ coupled with a controlled path $z_{k+j}$, with $j=0,\dots,m$. Let us define the regularized logarithm $L_\delta(t)$ as $\max\{\log|t|,\log\delta\}$. Note that $L_\delta(t)$ dominates $\log|t|$ for all $\delta>0$ and $L_\delta$ is $\delta^{-1}$-Lipschitz. We now have the chain of inequalities
\begin{align}
    \log R(z_{k+m})&=\log R(z_k)+\sum_{j=0}^{m-1}\log(2(1-X(z_{k+j})))
\\
&\le \log R(z_k)+\sum_{j=0}^{m-1}L_\delta(2(1-X(z_{k+j})))\\
&\le 
\log R(z_k)+4m\delta^{-1}\cdot 4^{-m} +\sum_{j=0}^{m-1}L_\delta(2(1-X(V_j)))\,.
\end{align}

On the one hand, to keep $z_{k+m}$ in control, we must have $\log R(z_{k+m})\ge Cm+\log R(z_k)$. Thus, the conditional probability in Eq.~\eqref{eq:condprob} is at most
\begin{equation}
    \mathbb{P}\bigg[\sum_{j=0}^{m-1}L_\delta(2(1-X(V_j))\ge Cm-4m\delta^{-1}\cdot 4^{-m}\ge Cm-\delta^{-1}\bigg]\,.
\end{equation}
On the other hand, we know the exact distribution of $V_m$: it is a uniform measure on the roots of order $2^m$ of $V$, i.e., a geometric progression with the step $e^{2\pi i\cdot 2^{-m}}$ on the circle that has one point on each arc of length $2\pi\cdot 2^{-m}$. Note now that if $|\zeta|=|\chi|=1$ and $|\zeta-\chi|<2\pi\cdot 2^{-m}$, then $|L_\delta(2(1-X(\zeta^{2^j})))-L_\delta(2(1-X(\chi^{2^j})|\le 2^{j-m} 2\pi\delta^{-1}$. Thus, if the sum $\sum_{j=1}^m L_\delta(2(1-X(\zeta^{2^j})))$ is above $Cm-\delta^{-1}$ anywhere on the arc, it is above $Cm-15\delta^{-1}$ on the whole arc.

It suffices to estimate the probability that
\begin{equation}
    \sum_{j=1}^m L_\delta(2(1-X(\zeta^{2^j}))\ge Cm-15\delta^{-1}
\end{equation}
where $\zeta$ is uniformly distributed on the unit circumference, which we will do utilizing some elementary Fourier analysis. Write
\begin{equation}
    L_\delta(2(1-X(\zeta)))=\sum_{q\in \mathbb Z} a_q\zeta^q\,.
\end{equation}
We then have $|a_0|\le \alpha(\delta)$ where $\alpha(\delta)\to 0$ as $\delta\to 0$ and $\sum_{q\ne 0}|a_q|\le \beta(\delta)<+\infty$ (explicit bounds are not hard, but we do not need them). Let $Z(\zeta)=\sum_{j=1}^m\zeta^{2^j}$. Then the sum we are interested in can be written as
\begin{equation}
\alpha(\delta)m+\sum_{q\ne 0}a_qZ(\zeta^q)\,.
\end{equation}
The first term can be made less than $\frac C2 m$ if $\delta>0$ is chosen small enough after $C>0$. So we are left with estimating the probability that
\begin{equation}\sum_{q\ne 0}a_qZ(\zeta^q)>\frac C2 m-15\delta^{-1}\,.\end{equation}
By the standard facts about lacunary series \cite{Zygmund2002}, we have $\|Z\|_{L^4}\le K_0\sqrt m.$ Hence the $L^4$ norm of the sum is at most $K_0\beta(\delta)\sqrt m$. Thus, the probability in question is bounded by
\begin{equation}
    \frac{K_0^4\beta(\delta)^2m^2}{(\frac C2m-15\delta^{-1})^4}\le \frac{A(C,\delta)}{n^2}
\end{equation}
when $n\ge N(C,\delta)$ (recall that $m\approx Cn$), which is better than we need if $n$ is also greater than $2A(C,\delta)$, say.
\end{proof}

    \section{Proof of complete unitary ergodicity in the Thue-Morse Drive}
\label{sec:prooftheorem}
In this section, we provide a proof of Theorem~1 on CUE on the single-qubit TMD.\footnote{We thank F. Nazarov for his help in this proof.} Lemma 2 in the main text follows from Proposition~\ref{prop:monotoneofdynamics} below. 

\begin{theorem}[Restatement of Theorem~1]
\label{theorem:CUEinTMD}
Consider a two-dimensional (qubit) system. For almost any $A,B\in \operatorname{SU}(2)$, the TMD satisfies CUE, $\mathcal{N}^{(k)}_\infty =\mathcal{N}^{(k)}_{\mathrm{Haar}}$ for all $k$, and consequently CHSE.
\end{theorem}
We will use the Hilbert-Schmidt or Frobenius norm on the superoperators of replicated space, \begin{equation}
    \norm{\mathcal{N}}_2=\sqrt{\sum_{\alpha \beta \gamma \delta} \abs{\bra{\alpha}\mathcal{N}\big[\dyad{\beta}{\gamma}\big]\ket{\delta}}^2},
\end{equation}
where the Greek indices label an orthonormal basis of the replicated Hilbert space $(\mathbb{C}^d)^{\otimes k}$. We will not use its explicit form: the key property to keep in mind is that it is unitarily invariant, $\norm{\mathcal{N}\circ \mathcal{V}}_2=\norm{\mathcal{N}}_2$ for any unitary channel $\mathcal{V}[\, \cdot \,]=V^{\otimes k}(\, \cdot\,)V^{\dagger \otimes k}$ with $V\in \mathrm{U}(d)$. Note also that, as $k$ will remain fixed throughout the proof, we are dropping the superindex $(k)$ on the channels over replicated space.

Recall that CUE establishes that the limit of the time-averaging channel is the Haar-averaging channel, $\lim_{T\to \infty} \mathcal{N}_T = \mathcal{N}_{\mathrm{Haar}}$. The first step in our proof is to realize that we only need to prove the convergence along the subsequence of exponential times $T=2^n$.
\begin{lemma}[Convergence along $T=2^n$ implies convergence] 
\label{lemm:convalongsubimpliesglobalconvergence}
The time-averaging channels converge
\begin{align}
    \lim_{T\to \infty} \mathcal{N}_T = \mathcal{N}_{\mathrm{Haar}} &&     \lim_{T\to \infty} \overline{\mathcal{N}}_T = \mathcal{N}_{\mathrm{Haar}}
\end{align}
if and only if they do so along the subsequence $T=2^n$, that is,
\begin{align}
    \lim_{n\to \infty} \mathcal{N}_{2^n} = \mathcal{N}_{\mathrm{Haar}} &&     \lim_{n\to \infty} \overline{\mathcal{N}}_{2^n} = \mathcal{N}_{\mathrm{Haar}}.\label{eq:convsubs}
\end{align}
\end{lemma}
\begin{proof}
Let us prove that $\lim_{T\to \infty} \mathcal{N}_T = \mathcal{N}_{\mathrm{Haar}}$ assuming Eqs.~\eqref{eq:convsubs}. The statement for $\overline{\mathcal{N}}_T$ is proven similarly. Let $\varepsilon>0$. From Eqs.~\eqref{eq:convsubs} there is some sufficiently large $n$ such that $\norm{\mathcal{N}_{2^n}-\mathcal{N}_{\mathrm{Haar}}}_2<\varepsilon/2$ and $\norm{\overline{\mathcal{N}}_{2^n}-\mathcal{N}_{\mathrm{Haar}}}_2<\varepsilon/2$. For every $m\in\mathbb{N}$, the corresponding segment $\omega_{2^n(m-1)+1}\omega_{2^n(m-1)+2}\cdots \omega_{2^nm}$ of the TMW coincides either with $\omega_{1}\omega_{2}\cdots \omega_{2^n}$ or with its bitwise negation. Hence, the time-averaging channel ${\mathcal{N}}_{[2^n(m-1):2^nm)}$ [see Eq.~\eqref{eq:partialaveraging}] is equal to either ${\mathcal{N}}_{2^n}\circ \mathcal{U}(2^n(m-1))$ or ${\overline{\mathcal{N}}}_{2^n}\circ \mathcal{U}(2^n(m-1))$, where $\mathcal{U}(t)$ is the unitary channel induced by the evolution operator $U(t)$, given by $\mathcal{U}(t)[\,\cdot\,]=U(t)^{\otimes k}(\, \cdot\,)U(t)^{\dagger \otimes k}$. We immediately get for $T=2^n N + r $, with $0\leq r<2^n$,

    \begin{align*}
\norm{\mathcal{N}_{T}-\mathcal{N}_{\mathrm{Haar}}}_2&\leq \frac{2^n }{T}\sum_{m=1}^N\norm{ {{\mathcal{N}}}_{[2^n(m-1):2^nm)}-\mathcal{N}_{\mathrm{Haar}}}_2+\frac{r}{T}\norm{{{\mathcal{N}}}_{[2^nN:2^nN+r)}-\mathcal{N}_{\mathrm{Haar}} }_2\\&\leq \frac{2^nN \varepsilon}{2T} + \frac{r}{T}2C\leq \frac{\varepsilon}{2} + \frac{2^n}{T} 2 C <\varepsilon,
    \end{align*}
    if we pick $T> 2^{n+2} C/\varepsilon$, where $C= \norm{{{\mathcal{V}}}}_2=\sqrt{d^{2k}}$ is the norm of any unitary channel $\mathcal{V}$. For the second inequality we used the unitary invariance of the norm and that $\norm{{{\mathcal{N}}}_{[a:b)}}_2,\norm{\mathcal{N}_{\mathrm{Haar}}}_2\leq C$.
\end{proof}

The rest of the proof will focus on showing that Eqs.~\eqref{eq:convsubs} are satisfied. To that end, we identify a positive quantity $M_n$ which is non-increasing in $n$ and is zero if and only if Eqs.~\eqref{eq:convsubs} are satisfied. We show that, in the qubit case $d=2$, whenever the disk dynamics recur, $M_n$ strictly decreases.

The following result implies Lemma 2 in the main text. 
\begin{prop}[Monotone of dynamics]
\label{prop:monotoneofdynamics}
    For each $n$, define
    \begin{align}
        M_n=\norm{\mathcal{N}_{2^n}-\mathcal{N}_{\mathrm{Haar}}}_2^2+\norm{\overline{\mathcal{N}}_{2^n}-\mathcal{N}_{\mathrm{Haar}}}_2^2.
    \end{align}
    This is $D_{2^n}^2 + \bar{D}_{2^n}^2$ in Lemma 2 in the main text. 
    Then $M_{n+1}\leq M_n$, and equality is attained if and only if ${\mathcal{N}}_{2^{n+1}}={\mathcal{N}}_{2^{n}}=\overline{\mathcal{N}}_{2^n}\circ \mathcal{A}_n$ and $\overline{\mathcal{N}}_{2^{n+1}}=\overline{\mathcal{N}}_{2^{n}}={\mathcal{N}}_{2^n}\circ \mathcal{B}_n$. 

        Moreover, if $K$ is a compact set in $\operatorname{SU}(d)\times \operatorname{SU}(d)$ such that each pair $(A,B)\in K$ generates $\operatorname{SU}(d)$, then there exist an open set $\Omega$ which contains $K$ and $\delta_* >0$ such that 
                \begin{equation}
                \label{eq:drop}
                M_{n+3}\leq (1-\delta_*)M_n
            \end{equation}
        if $(A_n,B_n)\in \Omega$. 
        
        Lastly, in the case of single qubit dynamics $d=2$, for almost all choices of starting unitaries $A,B\in \operatorname{SU}(2)$, there exists $\varepsilon$ such that for each $n$, if $\abs{c_n-c_1}<\varepsilon$ then Eq.~\eqref{eq:drop} holds, where $c_1$ is the disk coordinate corresponding to $\Phi(A,B)$ [this can be equivalently stated in terms of $(\xi_n,\theta_n)$ as in the main text].
\end{prop}
 Our main result follows straightforwardly from Proposition~\ref{prop:monotoneofdynamics}.  
 \begin{proof}[Proof of Theorem~\ref{theorem:CUEinTMD}]
 We know, by Lemma~\ref{lemm:lem01}, that the dynamics of the disk are recurrent, so $c_n$ comes back $\varepsilon$-close to $c_1$, infinitely often, where $\varepsilon$ is given by Proposition~\ref{prop:monotoneofdynamics}. Thus $\lim_{n\to \infty} M_n=0$, which implies, by Proposition~\ref{prop:monotoneofdynamics} that $\lim_{n\to \infty} \mathcal{N}_{2^n}=\lim_{n\to \infty} \mathcal{\overline{N}}_{2^n}=\mathcal{N}_{\textrm{Haar}}$. This yields CUE, by Lemma~\ref{lemm:convalongsubimpliesglobalconvergence}.
 \end{proof}
The proof of Proposition~\ref{prop:monotoneofdynamics} requires of the following technical result.
\begin{lemma} \label{lemm:M_monotone}Let $\mathcal{N}$ and $\mathcal{\overline{N}}$ be any quantum channels in the $k$-replicated Hilbert space and $A,B\in\operatorname{SU}(d)$. Define the following dynamical map
 \begin{align}
        \label{eq:Gdef}G(\mathcal{N},\overline{\mathcal{N}},A,B)=\left(\tfrac{1}{2}(\mathcal{N}+\overline{\mathcal{N}}\circ \mathcal{A}),\tfrac{1}{2}(\overline{\mathcal{N}}+\mathcal{N}\circ \mathcal{B}),BA,AB \right),
    \end{align}
    where $\mathcal{A}[\,\cdot\,]=A^{\otimes k}[\,\cdot\,]A^{\dagger \otimes k}$ and $\mathcal{B} [\,\cdot\,]=B^{\otimes k}[\,\cdot\,]B^{\dagger \otimes k}$.
    Let $M(\mathcal{N},\overline{\mathcal{N}})=\norm{\mathcal{N}-\mathcal{N}_{\mathrm{Haar}}}_2^2+\norm{\overline{\mathcal{N}}-\mathcal{N}_{\mathrm{Haar}}}_2^2$. Then $M(\mathcal{N},\overline{\mathcal{N}})\leq M(G(\mathcal{N},\overline{\mathcal{N}},A,B))$, \footnote{$M(G(\mathcal{N},\overline{\mathcal{N}},A,B))$ means we evaluate $M$ on the first two entries of $G(\mathcal{N},\overline{\mathcal{N}},A,B)$.} and equality is attained if and only if $\mathcal{N}=\overline{\mathcal{N}}\circ \mathcal{A}$ and $\overline{\mathcal{N}}={\mathcal{N}}\circ \mathcal{B}$.

    Moreover, if $K$ is a compact set in $\operatorname{SU}(d)\times \operatorname{SU}(d)$ such that each pair $(A,B)\in K$ generates $\operatorname{SU}(d)$, then there exist an open set $\Omega$ which contains $K$ and $\delta_* >0$ such that for any $(A,B)\in \Omega$ and channels $\mathcal{N},\overline{\mathcal{N}}$ \begin{equation}
        M(G^3(\mathcal{N},\overline{\mathcal{N}},A,B))\leq M(\mathcal{N},\overline{\mathcal{N}})(1-\delta_*).
    \end{equation}
\end{lemma}
\begin{proof}[Proof of Lemma~\ref{lemm:M_monotone}]

Let us denote for brevity $\mathcal{D}=\mathcal{N}-\mathcal{N}_{\mathrm{Haar}}$ and $\overline{\mathcal{D}}_{n}=\overline{\mathcal{N}}-\mathcal{N}_{\mathrm{Haar}}$.
We have

\begin{align*}
   M(G(\mathcal{N},\overline{\mathcal{N}},A,B))&=\frac{1}{4}\norm{\mathcal{D}+\overline{\mathcal{D}}\circ \mathcal{A}}_2^2+\frac{1}{4}\norm{\overline{\mathcal{D}}+\mathcal{D}\circ \mathcal{B}}_2^2  
   \\ &\leq \frac{1}{4}\left(\norm{\mathcal{D}}_2+\norm{\overline{\mathcal{D}}\circ \mathcal{A}}_2\right)^2+ \frac{1}{4}\left(\norm{\overline{\mathcal{D}}}_2+\norm{\mathcal{D}\circ \mathcal{B}}_2\right)^2
   \\ &= \frac{1}{2}\left(\norm{\mathcal{D}}_2+\norm{\overline{\mathcal{D}}}_2\right)^2
   \\ &\leq \norm{\mathcal{D}}_2^2+\norm{\overline{\mathcal{D}}}_2^2 = M(\mathcal{N},\overline{\mathcal{N}}),
\end{align*}
where we used the unitary invariance of the norm in the third equality.
The first inequality is a triangle inequality, which attains equality if and only if $\mathcal{D} = a\overline{\mathcal{D}} \circ \mathcal{A}$ and $\overline{\mathcal{D}} = b\mathcal{D} \circ \mathcal{B}$ for some $a,b\geq 0$. Additionally, the last inequality is an equality if and only if $\norm{\mathcal{D}}_2=\norm{\overline{\mathcal{D}}}_2$, which implies $a=b=1$.

Now, for the second part, let us  assume that the pair $(A,B)$ generates $\operatorname{SU}(d)$ and denote $\mathcal{N}',$ $\overline{\mathcal{N}'}$ the channels obtained by applying the dynamics \eqref{eq:Gdef} three times. Observe that if $M(\mathcal{N},\mathcal{N})=M(\mathcal{N}',\overline{\mathcal{N}'})$, then by the previous part \begin{align*}
    \mathcal{N}=\overline{\mathcal{N}}\circ \mathcal{A}=\overline{\mathcal{N}}\circ \mathcal{B}\mathcal{A}=\overline{\mathcal{N}}\circ \mathcal{A}\mathcal{B}\mathcal{B}\mathcal{A}  &&  \overline{\mathcal{\mathcal{N}}}=\mathcal{N}\circ \mathcal{B} =\mathcal{N}\circ \mathcal{A}\mathcal{B}=\mathcal{N}\circ \mathcal{B}\mathcal{A}\mathcal{A}\mathcal{B}, 
\end{align*}
which imply that $\mathcal{N}\circ \mathcal{A}=\mathcal{N}\circ \mathcal{B}=\mathcal{N}$ and $\overline{\mathcal{N}}\circ \mathcal{A}=\overline{\mathcal{N}}\circ \mathcal{B}=\overline{\mathcal{N}}$, which further implies  $\mathcal{N}= \mathcal{N}_{\textrm{Haar}}$ and $\overline{\mathcal{N}}= \mathcal{N}_{\textrm{Haar}}$ because $(A,B)$ generates $\operatorname{SU}(d)$. In this case, $M(\mathcal{N},\overline{\mathcal{N}})=0$. Thus, we have shown that $\delta(\mathcal{N},\overline{\mathcal{N}},A,B)=1- M(\mathcal{N}',\overline{\mathcal{N}'})/M(\mathcal{N},\overline{\mathcal{N}})>0$ whenever $M(\mathcal{N},\overline{\mathcal{N}})>0$ and $(A,B)$ generates $\operatorname{SU}(d)$. Let $\delta_*$ be the minimum of $\delta$ over all $(A,B)\in K$ and superoperators $\mathcal{N},\mathcal{N}'$ such that $M(\mathcal{N},\overline{\mathcal{N}})=1$. Then $\delta_*>0$ because we optimized a continuous function over a compact set. Now, for pairs for which  $1\neq M(\mathcal{N},\overline{\mathcal{N}})>0$, note that $\delta$ is invariant upon the affine transformation $(\mathcal{N},\overline{\mathcal{N}})\mapsto (\lambda\mathcal{N}+ (1-\lambda)\mathcal{N}_{\textrm{Haar}},\lambda\overline{\mathcal{N}}+ (1-\lambda)\mathcal{N}_{\textrm{Haar}})$, with $\lambda>0$. As one can always ensure $M(\mathcal{N},\overline{\mathcal{N}})=1$ by picking an appropriate $\lambda$, we have $\delta\geq\delta_*>0$ over all pairs $(\mathcal{N},\overline{\mathcal{N}})$ such that $M(\mathcal{N},\overline{\mathcal{N}})>0$. Because $\delta\geq\delta_*>0$ holds over the compact set $K$ and $\delta$ is continuous, we can guarantee that $\delta\geq\delta_*/2>0$ holds over an open set $\Omega$ which contains $K$. The claim follows by relabeling $\delta_* \to \delta_*/2$.
\end{proof}

\begin{proof}[Proof of Proposition~\ref{prop:monotoneofdynamics}]
    Note that precisely $(\mathcal{N}_{2^{n+1}},\overline{\mathcal{N}}_{2^{n+1}},A_{n+1},B_{n+1})=G(\mathcal{N}_{2^n},\overline{\mathcal{N}}_{2^n},A_n,B_n)$, with $G$ defined as in Eq.~\eqref{eq:Gdef}, and $M_n=M(\mathcal{N}_{2^{n}},\overline{\mathcal{N}}_{2^{n}})$, so the first two claims for general $d$ follow immediately.  For the last claim on $d=2$, let $K$ be the set of all pairs $(A_*,B_*)$ with the same eigenvalues and whose corresponding disk coordinate is $c$. It is straightforward to see that with unit probability over the choice of $(A,B)$, all pairs in $K$ generate $\operatorname{SU}(2)$, because generically their common angle of rotation is an irrational multiple of $\pi$ and their rotation axes are not parallel. By Lemma~\ref{lemm:M_monotone} there is an open set $\Omega\supseteq K$ and $\delta_*>0$ such that $M_{n+3}\leq (1-\delta_*)M_n$ if $(A_n,B_n)\in \Omega$. The claim follows by picking $\varepsilon>0$ small enough so that $(A_n,B_n)\in \Omega$ if $\abs{c_n-c_1}<\varepsilon$.
\end{proof}
\bibliography{references}